# ACCEPTANCE OF E-PROCUREMENT IN ORGANISATIONS: USING STRUCTURAL EQUATION MODELING (SEM)

Muhammed S. Maddi[1], Paul Davis[2] and John Geraghty[3]

[1]Dublin City University, Glasnevin, Dublin 9, Ireland & College of economic and Management Bani Walid City University, Libya
[2]Dublin City University Business School, Glasnevin, Dublin 9, Ireland
[3]Dublin City University Mechanical and Manufacturing Engineering School, Glasnevin, Dublin 9, Ireland

## ABSTRACT

*This research is concerned with the development of a realistic model for e-procurement adoption by organizations and groups observing the Rules of Islamic Sharia (RIS). This model is intended to be based on the behavioural control, subjective norms, and the recognition of the benefits and risks of e-procurement adoption. The developed model, "E-Procurement Adoption Model" (E-PAM), combined and extended two existing models previously used for information technology adoption. Central to the design of the E-PAM is the principle that a realistic model should consider all relevant psychological, social, cultural, demography, and religious factors. The mediating components "mediators" that determine the final model are thus determined by the specific aforementioned factors about of on to the development of an e-procurement approach in the context of the Rules of Islamic Sharia. Therefore, in the design of the research model, factors including perceived usefulness, perceived ease of use, organizational facilitators, organisational leadership, critical success factors, challenges, barriers, and benefits were taken into account. These factors in turn gave rise to the two mediating components of 'Attitude' and 'Rules of Islamic Sharia'.*

*Several hypotheses were made in order to establish links between the contributing factors, the mediators, and the adoption of e-procurement as the final dependent variable output of the model. An online questionnaire survey was conducted to validate the hypotheses using a structural equation model. The descriptive analysis of the survey data provided demographic details of the questionnaire participants and their employers and statistical analysis methods were used to correlate the contributing factors and mediators. The analysis of the survey data confirmed the existence of positive relationships between Benefits and the Rules of Islamic Sharia, organizational Facilitators and the Adoption of e-procurement, Organisational leadership, and Rules of Islamic Sharia and the Adoption of e-procurement. It was observed that Perceived Usefulness and Organisational Leadership do not influence the adoption of e-procurement. Furthermore, the analysis suggested that Rules of Islamic Sharia is not influenced by Critical Success Factors. This model successfully validated most of the initial hypotheses made regarding e-procurement adoption. The results of this research strongly encourage organisations observing Rules of Islamic Sharia to adopt e-procurement practices*

# 1. INTRODUCTION

The adoption of new technology, or the acceptance of a new innovative system for traditional business management processes, depends upon the demographic and psychological characteristics of the target audience. The adoption of this new technology can be achieved efficiently by developing and utilizing realistic models [28], which are commonly referred to as Technology Acceptance Models (TAM). The contributing factors of such models should include important psychological, social, cultural and demographic characteristics, which have a tremendous influence over the final output of a model. If the intended groups belong to a certain religion and engage religious practices in business, it is important to include these religious provisions in the TAMs that influence the overall adoption process. In this research, a new model is developed and utilised to encourage the organisations working in compliance with Islamic provisions to accept and adopt e-procurement systems.

The selection of the relevant contributing factors for a TAM depends upon the traditional methods and routines which have been typically used within organizations where new technologies are required to be put in place. Before developing a new model for technology acceptance, it is important to understand the existing business models, the contributing factors which influence the overall process, additional factors (such as cultural or religious, etc.) that could influence the process, the new technology that is to be adopted, and the advantages and disadvantages of the new technology.

The Procurement process influences the overall performance of the development and growth of any industry. Most of the existing studies on the procurement process focus only on specific geopolitical regions such as North America, Central, and Western Europe, and Asia. The lack of research relating to e-procurement in the regions of North Africa and the Middle East is evident from a survey of the existing literature. Furthermore, studies in this field have been historically performed by focusing on the buyer side of e-procurement, thus ignoring the supplier side. More recently, research studies aimed at encouraging and promoting the adoption of e-procurement have begun to focus on both buyers and suppliers. However, there remains a lack of empirical studies on how a recognition of the benefits of e-procurement (BNF), challenges and barriers (CHB), critical success factors (CSF) and influence of Rules of Islamic Sharia might help business and research communities obtain a deeper intellectual understanding of e-procurement implementation in various economic world regions by considering differences of culture, religion, politics, and tradition. The outcomes of this research project will provide senior management in purchasing and procurement departments with new insights into the importance of e-procurement from the management point of view and will also provide important details on the operation of this technology in many developed countries. The results of this research should encourage small and medium- sized enterprises (SMEs) to adopt e-procurement as an innovative way of maintaining relationships with their suppliers and clients.

E-procurement has been used quite extensively by organizations in both the public and private sectors as it provides smart, efficient strategies and solutions for sustainable growth and competitive advantage for organizations globally [44]. To understand e-procurement systems, it is important to first understand the basic principles and processes that have been used in traditional procurement systems. In the following sections the basis of procurement, specifically in relation to supply chain management, and other important procurement processes are discussed.

## 2. TECHNOLOGY ADOPTION: THEORETICAL OVERVIEW

The pace at which new technology is accepted and implemented determines its success and impact on society. Organizations currently live in a global village or marketplace where it is important to determine the success and impact factors of a new technology across the world. Adopting a new technology is a complex process that involves not only different customers but also the innovators who adopt a new technology within an Organization. The innovators within an Organization are the group of people who are willing to take risks and that possess business acumen. The customers of a new technology could be divided into different parts with respect to the time duration it takes for them to adopt a new technology. The pace of technology adoption is dependent upon several economic/financial, religious, cultural and social factors. The relationship between technology and acceptance can also be affected by contemporary methods of communication and new media such as Google and Twitter [21]. The advent of the industrial revolution and rapid development of new technologies and innovations in existing systems poses a continual requirement for useable models and frameworks, based upon the aforementioned factors, to convince the customers to adopt and implement such new systems. Such models are designed on the basis of important factors contributing towards the decision to adopt and implement a new technology. The factors in a model are often mediated by dependent variables and the link between factors and mediators could be established by making certain hypotheses. The successful execution of (and receiving expected results from) the model validates the hypotheses. The validation of such hypotheses provides guidelines to innovators and technology developers to convince the customers about adopting new technology. In addition to this, research studies based on such models encourage customers, who belong to certain social, cultural and religious background, to adopt a particular technology [21]. The model developed in this research, E-PAM, is designed around contributing factors such as usefulness, organisation facilitators and organisation leadership and the mediators of attitude and RIS. Attitude was used in a model presented previously, TAM, regarding technology adoption [26]. A new mediator was used in E-PAM to describe the involvement rules of RIS in adoption of e-procurement. RIS is a cultural factor which was not studied previously and, one of the main concepts belonging to RIS relates to the morality of dealing with trade and business. In this section, a number of models from literature related to adoption, acceptance of new and innovative technologies were discussed. The contributing factors and mediators used in E-PAM in accordance to their use in models present in literature were highlighted. [26],[29].

### 2.1. Rules of Islamic Sharia (RIS)

Rules of Islamic Sharia offer special provisions, rules and regulations defining concepts, and the operational processes of organizations. These provisions, rules and regulations of Islamic Sharia prohibit any kind of transactions involving betting a sum of money. Furthermore, the Sharia rules prohibit any transactions leading to earning interests on loaned money. Interest in Islamic terms is referred to as Riba (usury). Riba signifies the stipulated excess over and above the principal that a loan is required to pay to the lender [24]. Any charge for the privilege of borrowing money is essentially forbidden in Islamic Sharia [11].

Hence, the main question that arises concerns the way entities make money under this particular law. The banks following RIS can earn profit mainly by sharing the risks with their consumers and then dividing the profit. The main items or investment options permitted or defined under the Islamic laws are Ijara (rental), Ijara-wa-iqtina (rental and possession), Mudaraba (speculation), Murabaha (sharing in profits) and Musharaka (partnership). While in Ijara the bank makes a purchase and then leases it back, under Mudaraba financial experts are hired to help in generating profits that are then shared [53].

All these laws are based on one basic factor which is the lack of importance given to money. Money is simply considered to be a medium of exchange that has no intrinsic value. Hence, there are some specific ways in which the companies can share their risks with the Islamic banks and earn profits. They can enter into an agreement called Musharaka, wherein the banks and the owners invest equally into a venture and the returns and profits are then shared among them [1].

Islamic provisions define and require the people to be extremely productive but also lawful and hence not only do the businesses to earn money, but they also do it in a way that is extremely fair to society. The Islamic laws promote trading and business activities as highly important, and allow for the concept of the e-commerce to be developed and implemented while observing RIS [46]. The importance and significance of the e-commerce is self-evident because it facilitates accuracy, flexibility, convenience and speed. The RIS provides guides to traders and businessmen on how to perform business ethically and avoid deception [50]. Therefore, by including RIS as a mediator in an acceptance and adoption model for e-procurement could help individuals and organizations observing RIS use state of the art e-procurement systems.

## 2.2. Electronic Procurement Adoption Model (E-PAM)

The developed model, "E-Procurement Adoption Model" (E-PAM), combined and extended two existing models previously used for information technology adoption. Central to the design of the E-PAM is the principle that a realistic model should consider all relevant psychological, social, cultural, demographic and religious factors.

For the development of a model for new technology adoption, it is important to understand the factors that contribute to the paradigm of reluctance towards certain technology, as these factors contribute either positively or negatively towards the final output of a model. The characteristics of a model could be established by selecting the appropriate contributing factors. The models described above are mainly distinct due to the selection of particular contributing factors while discounting certain others. The selection of contributing factors is mainly based on the degree of influence that they could provide over a certain technology and socio-economic conditions. The contributing factors selected in the design and development of E-PAM model in this research are discussed below with relevant use in previous models.

## 2.3. Research Hypotheses

According to TAM, perceived usefulness affects a person's attitude towards using a system. Lai and Yang argued that employees in a performance-oriented e-business context are generally reinforced for good performance and benefits [27]. This implies that realizing the usefulness of e-business applications, such as mobile banking in improving performance or efficiency, will positively impact attitude towards that application. The effect of perceived usefulness on attitude has been validated in many studies including [29][19], [48], [25].

Consequently, the following hypotheses H1and H5 are suggested:

**Hypothesis 1:** Perceived usefulness (PU) has a direct, positive effect on an organization's desire to adopt e-procurement (AEP).

Many existing studies in the context of e-business have shown that an individual's attitude directly and significantly influences behavioral intention to use a particular e-business application [35], [15] [16]. For example, George, J, found a strong positive relationship between an individual's attitude toward purchasing online and the user's behavioural intention [15]. Gribbins etl., studied the acceptance of wireless technologies by users [16]. Also Püschel J. etl. founded that attitude significantly affects intention to adopt mobile banking. They demonstrated support for the relationship between attitudes toward using mobile commerce/ banking and behavioral intention [17]. Thus, the following hypotheses are proposed:

**Hypothesis 2:** Attitude positively influences the e-procurement adoption.

**Hypothesis 3:** Organisation facilitators will have a positive effect on e-procurement adoption.

**Hypothesis 4:** Organisation Leadership will have an immediate positive effect on e-procurement adoption.

**Hypothesis 5:** Perceived usefulness positively influences attitudes towards e-procurement adoption.

Previous studies on PEOU provide empirical evidence supporting its impact on attitude and usefulness. Examples include mobile banking in Malaysia, internet banking acceptance, wireless finance, and mobile commerce [30],[51],[30] [31], [25],[5],[39]. Users would be concerned with the effort required to use that application and the complexity of the process involved. Such perceived ease of browsing, identifying information and performing transactions should enable favorable and compelling individual experience [42], [38]. TAM suggests that ease of use is thought to influence the perceived usefulness of a technology. The easier it is to use technology, the greater the expected benefits from the technology with regard to performance enhancement. This relationship has also been validated in online technology context [13][14] [33][34] [36]. Thus, this research examines the following hypotheses:

**Hypothesis 6:** Perceived ease of use positively influences attitudes towards e-procurement adoption.

**Hypothesis 7:** Perceived ease of use positively influences the perceived usefulness of e-procurement adoption.

The above hypotheses provide an interesting relationship between perceived usefulness and ease of usage of new technology and their link to the adoption of the technology itself. Davis, [10] demonstrated that the perception of ease of use and usefulness of technology are significant influences affecting the attitude of people and it, in turn defines the acceptance or adoption of e-procurement. This has been proved by further studies conducted using Davis's TAM,[29], [48], [50],[37].

Kaliannan, etc. demonstrates that the two factors PU and EOU, as well as the organization leadership and organization facilitators, are also important factors for e-procurement adoption in organisations [23]. Hence, the first five hypotheses propose that irrespective of culture and other factors, people will develop an attitude to adopt e-procurement only when they see that it is useful. However, the social conditions and cultural factors of a society contribute significantly towards the acceptance of a new technology [45], [3]. Therefore, hypotheses H3, H4, H8, and H9 focus on the organizations cultural beliefs and value systems, and they are represented in this research as follows:

**Hypothesis 8:** Organisation facilitators will have a positive effect on the perceived ease of use.

**Hypothesis 9:** Organisation facilitators will have a positive effect on the organization's leadership.

The RIS mediator was placed in the model to moderate the effect of benefits, critical success factors, and barriers and challenges of e-procurement adoption. The mediator variable in any model is responsible for explaining and establishing a link between dependent and independent variables. In E-PAM, the independent variables provide input to the model and influence the final output (AEP). This relationship between input and output in E-PAM was mediated by two variables, namely "Attitude" and "Rules of Islamic Sharia (RIS)". The attitude has been widely discussed and used by researchers for technology adoption and implementation (TAM, TPB, etc.). As noted from literature, RIS has never been used as a mediator to establish a link between independent and dependent variables for e-procurement technology adoption and acceptance. A number of independent factors or variables influence the e-procurement adoption, and in order to keep consistency with previous research works, three variables (CSF, CHB, and BNF) have been selected to be mediated by RIS in E-PAM. These three variables have been used before in e-procurement technology acceptance model [18]. For mediation models (models with at least one mediator), the relationship between independent variables and mediator and the relationship between the mediator and dependent variable should be examined. Such examination could provide mediation information about the intervention caused by the mediator. In this way, the role of a mediator (positive or negative) could be established in a model. The mediation caused by the mediator could be investigated by finding the correlation between the independent variables and the mediator. The correlation between the independent and mediator (termed as collinearity) describes the nature of mediation caused by the mediator. The nature of the mediation could be supporting (encouraging) or not supporting (discouraging) between the independent and dependent variables.

In order to investigate the above-mentioned relationships between the mediator and independent and dependent variables, three hypotheses were formulated. These hypotheses were made to provide a complete picture of the intervention caused by the mediator (RIS) by establishing the links between CSF, CHB, and BNF and AEP through RIS. The three hypotheses are as follows:

**Hypothesis 10:** Rules of Islamic sharia will have a positive effect on the employees to adopt e-procurement technology.

**Hypothesis 11:** Critical success factors will have a positive effect on the employees toward e-procurement technology.

**Hypothesis 12:** Challenges and barriers will have a positive effect on the employees toward e-procurement technology.

**H13:** Benefits will have a positive effect on the employees toward e-procurement technology [28].

The successful proof of these hypotheses could therefore predict the e-procurement technology adoption guidelines for RIS observing users and organizations. The contributing factors that are correlated by the hypotheses made to develop the E-PAM are present in Figure 1.

## 2.4. Data Collection

In this study, quantitative research methods have been used to investigate the possibilities of e-procurement acceptance and adoption by users and organizations observing the Rules of Islamic Sharia (RIS). For this purpose, previously developed technology acceptance models were used to construct a new model based on a cultural mediator, i.e., RIS, which has never been studied before for e-procurement acceptance and adoption. In this study, quantitative research methods have been used to investigate the possibilities of e-procurement acceptance and adoption by users and organisations observing Rules of Islamic Sharia (RIS). For this purpose, previously developed technology acceptance models were used to construct a new model based on a cultural mediator, i.e., RIS, which has never been studied before for e-procurement acceptance and adoption. The data from completed questionnaires were collected using Survey Monkey®. The data were exported to Microsoft Excel sheets from Survey Monkey®.

The data collected from the survey was processed and analyzed systematically using the quantitative methods which are appropriate for such research studies [6]. The collected data were stored in two different groups. The first group of data was related to the information about the employees (survey participants) and a brief description about their company background. These descriptive data were processed for descriptive analysis. The statistical data collected from the respondents was statistically analysed using appropriate commercially available statistical SPSS software.

The structure of the survey is presented in Appendix A. The online survey contained 23 questions which were grouped into three sections. The first section contained nine questions (Q1-Q9) that were designed to collect information to develop a profile of the respondents and their organizations. The second section consisted of ten questions (Q10-Q19) that were designed to elicit information to test the research hypotheses and the validity of the proposed E-PAM model. The third section, which consisted of five (Q20-Q24), asks about the e-procurement tools and the length of time for which they have been in use at the organisation, in order to estimate their experience and expertise. It asks for the e-procurement tools to be ranked based on their levels of benefits in achieving company objectives. Respondents were also asked to provide optional comments on the survey and themselves.

## 2.5. Sample

Email carrying online questionnaire link has been sent to 550 randomly selected organizations from eleven countries around the world four of them categorized as developed countries, and the rest are developing countries. The most of responses were from Ireland, Libya, UK, and USA. The online questionnaire has more advantages than an offline survey. The reason for that it makes data collection and analysis, to let the organisations key positions to answer them in a limited period and to use the online survey technique to the analysis the results faster and easier. More details about the data collections were presented in chapter three at section 3.8 of the research methodology.

## 2.6. Methodology

Structural Equation Modelling SEM is an analytical approach in which one can perform factor analysis and combine it with the linear regression models for validating the hypotheses of a theory. SEM is one of the most popular data analytic techniques for testing and validating survey data [52]. Many researchers in management and information technology fields use SEM and they have demonstrated that it is an appropriate tool for testing theories and combing the analysing of factors with linear regression [8]. SEM is a multivariate method that was used to assess the reliability and validity of the model measures. It was designed to test the theoretical framework or conceptual model of the study. Common SEM methods include discriminant validity, confirmatory factor analysis and latent growth modelling [2], [40],[41]. The Discriminant validity method is used to validate the independency of concepts that are supposed to be dependent upon each other. Confirmatory factor analysis procedure provides information about the accuracy of measured variables representing the number of constructs, while latent growth modeling is another statistical method commonly used to estimate growth trajectory.

There are several advantages of using SEM in analyzing data collected from the survey. These are as follows:

1. SEM allows hidden variable models and isolated approximation of the relationship between hidden factors and corresponding indicators, as well as between the factors themselves.

2. Using SEM, the measurement of global fit to collected data can be provided for complex models that are based upon the huge number of equations.

3. SEM is a good alternative to straw-man testing of a model that either reject or fail to reject the null hypothesis.

The graphical diagram of Structure Equation a Model SEM representing the E-PAM could be drawn using AMOS. AMOS is a special software package for SPSS to perform structural equation modelling. The following steps were performed to draw the SEM on AMOS and upload the survey data obtained by online survey model.

1. SEM graphical designing using AMOS: the AMOS graphical user interface was used to draw the graphical diagram of SEM.
2. Reading data into AMOS: external data could be uploaded to the SEM using AMOS. The online survey data was exported to EXCEL sheets by Survey Monkey. The data from EXCEL sheets were read by AMOS and delivered to SEM graphical diagram.

3. Selecting analysis properties: the appropriate analysis properties were selected including a standardized solution, squared multiple correlations, the sample covariance matrix, and the covariance matrix of the residuals.

4. Model running and analysis: once the data was fed to SEM model, the model was run through AMOS.

## 3. DATA ANALYSES RESULTS

Before to the analysis, the data were examined to ensure that there are no outliers present. The dataset was observed for missing values, which averaged 4.7% of the cases, and replaced with values obtained by the expectation-maximization (EM) iterative algorithm implemented in SPSS version 21.0. The algorithm estimates parameters for missing value by the maximum likelihood method. The (EM) method is seen to be a more accurate algorithm than other substitution and elimination techniques [22].

## 3.1. Reliability and Unidimensionality Test

The test for reliability was conducted by Chronbach's alpha estimation. The results of this test are presented in Table 1. Chronbach alpha estimation is a common test to measure the internal consistency or reliability that indicates how well a set of variables measure a single or one-dimensional construct. The cut-off value for Chronbach alpha was considered as 0.7 in this study, above which concerned variables reliably measure the underlying construct. Chronbach alpha of most constructs was above 0.7, and the same measure of the construct, ATT, was 0.5. This observation indicated that the underlying items reliably measured the corresponding construct. Table 1 also presents the first Eigenvalue and the percentage variance explained by each construct. The results showed that all constructs except ATT and AEP received Eigenvalue greater than one. This indicated uni-dimensionality of the construct. Therefore, the variables measured a single dimension of the construct as it was modeled theoretically. It was also observed that all constructs explained variance well above 50% that further established the notion of uni-dimensionality.

Table 1. Chronbach's alpha, Eigenvalue and variance

| Construct | Chronbach's Variance | alpha explained | Eigenvalue |
|-----------|----------------------|-----------------|------------|
| ATT | 0.500 | 0.530 | 63.304% |
| PU | 0.909 | 3.145 | 78.620% |
| OF | 0.830 | 3.074 | 61.470% |
| OL | 0.903 | 3.690 | 72.575% |
| EOU | 0.715 | 1.918 | 63.940% |
| RIS | 0.887 | 3.963 | 66.046% |
| CHB | 0.910 | 3.675 | 73.499% |
| CSF | 0.955 | 7.602 | 69.105% |
| AEP | 0.921 | 0.592 | 86.395% |
| BNF | 0.970 | 9.610 | 73.921% |

55% of respondents of the online questionnaire survey hold important positions such as director, CEO, manager, or engineer. The rest of the respondents hold positions of procurement manager, sales/marketing manager, consultant, or administrator, etc. Most of these respondents were working in the areas of engineering, suppliers, contracting/tendering, or general management, etc. Around 73% of respondents had more than 5 years of experience, 29.1% of them had more than 15 years of experience while 25.3% of respondents had 5 or fewer years of experience. The description of the respondents indicated that the survey response came from well-experienced professionals. Around 78% of respondents had bachelors or higher degree indicating their high academic qualifications.

28.5% respondents belong to firms having 500 or fewer employees. A Similar percentage of respondents belongs to firms having more than 2500 employees. The firm sizes of remaining respondents were between 500 and 2500 employees.

## 3.2. Correlation Coefficient and Multicollinearity

Appendix (B) shows the inter item Pearson correlation coefficients. These coefficients are low to moderate in magnitude. If these coefficients exceed 0.9 then the possibility of multicollinearity being present in the data is high [20], The Problem of multicollinearity arises when correlations among predictors are high which makes results unstable and difficult to interpret. Since no correlation coefficient exceeds 0.9, the problem of multicollinearity did not appear to be present. Confirmatory factor analysis (CFA) based structural equation modeling is used to measure and test the conceptual model using survey data. The statistical program AMOS version 21.0 was used to test the measurement and structural model and SPSS version 21.0 was used for descriptive analyze. As per [54] suggestions, two-step approach-measurement model and structural model has been used to analyse the model with collected data.
.
Table 2 presents all constructs and the respective measurement items and loadings. The loadings describe how much a variable contributes in explaining its underlying factor.

### 3.3. Convergent Validity

Convergent validity used to measure the similarity between the individual items used to measure the same construct. This measure gives a clue that individual items are well related, and they together measure the underlying construct. Convergent validity was assessed using standardized parameter loading of the measurement items on their respective construct. Items that did not load significantly, i.e., if loading is less than 0.50, those constructs were removed from the model. Convergent validity was assessed using standardised parameter loading of the measurement items on their respective construct. All the loadings ranged from 0.67 and 0.90 are significant (p-value<0.01) providing support for convergent validity. The Average variance extracted (AVE) and composite reliability (CR) is also computed as suggested by [12]. The AVE and CR are presented in Table 5.1. The CR ranged from 0.71 to 0.97, and exceeded the suggested cut-off of 0.7, AVE ranged from 0.74 to 0.88, exceeding the cut-off of 0.5. This conveys that variance captured by a factor is more than the variance of error component.

Table 2. Constructs and their item loading

| Constructs | (AVE) | (CR) | Items | Item loading |
|---|---|---|---|---|
| (PU) | 0.84 | 0.90 | PU1 | .859 |
| | | | PU2 | .844 |
| | | | PU3 | .828 |
| | | | PU4 | .848 |
| (A) | 0.78 | 0.87 | ATT1 | .854 |
| | | | ATT2 | .913 |
| | | | ATT3 | -0.314 |
| | | | ATT4 | .738 |
| (PEOU) | 0.74 | 0.71 | PEOU1 | .498 |
| | | | PEOU2 | .674 |
| | | | PEOU3 | .816 |
| Organizational facilitators | 0.75 | 0.84 | OF1 | .713 |
| | | | OF2 | .861 |
| | | | OF3 | .750 |
| | | | OF4 | .683 |
| | | | OF5 | .045 |
| Organisational leadership | 0.80 | 0.90 | OL1 | .792 |
| | | | OL2 | .795 |
| | | | OL3 | .781 |
| | | | OL4 | .813 |
| | | | OL5 | .842 |
| Rules of Islamic Sharia | 0.86 | 0.93 | RIS1 | .890 |
| | | | RIS2 | .895 |

| Constructs | (AVE) | (CR) | Items | Item loading |
|---|---|---|---|---|
| | | | RIS3 | .767 |
| | | | RIS4 | .851 |
| | | | RIS5 | .897 |
| Benefits (BNF) | 0.85 | 0.97 | BNF1 | .842 |
| | | | BNF2 | .868 |
| | | | BNF3 | .844 |
| | | | BNF4 | .891 |
| | | | BNF5 | .859 |
| | | | BNF6 | .877 |
| | | | BNF7 | 0.205 |
| | | | BNF8 | 0.081 |
| | | | BNF9 | 0.055 |
| | | | BNF10 | 0.278 |
| | | | BNF11 | 0.070 |
| | | | BNF12 | 0.019 |
| | | | BNF13 | 0.017 |
| Challenges and Barriers | 0.85 | 0.90 | CHB1 | .823 |
| | | | CHB2 | .809 |
| | | | CHB3 | .809 |
| | | | CHB4 | .832 |
| | | | CHB5 | .816 |
| Critical success factors | 0.81 | 0.95 | CSF1 | .772 |
| | | | CSF2 | .823 |
| | | | CSF3 | .859 |
| | | | CSF4 | .836 |
| | | | CSF5 | .821 |
| Adoption of e-procurement | 0.88 | 0.91 | AEP1 | .895 |
| | | | AEP2 | .851 |
| | | | AEP3 | .906 |

# 4. DESCRIPTIVE STATISTICS

## 4.1. Organisations' Country

E-procurement is largely dependent on the extent of the technology use, and this use of technology and its perceived usefulness vary across different countries. The 60 respondents (32.3%) were from Ireland, 36 respondents (19.4%) were from Libya, 21 respondents (11.3%) were from the U.K., 15 respondents (8.1%) were from the U.S., 10 respondents (5.4%) were from Malaysia, 9 respondents (4.8%) were from Saudi Arabia, and 8 respondents (4.3%) were from the UAE. The 51.7% of responses were from Ireland and the U.K. It is also worth noting that 31.2% of the respondents were from Libya, UAE, Saudi Arabia, and Kuwait and 12.7% were also from other different Arab countries. Therefore, the total of the responses from Islamic and Arab countries was almost 40%.

Table 3 Organisations' Country of Operations

| Country | Frequency | Percent |
| --- | --- | --- |
| Kuwait | 5 | 2.7 % |
| Egypt | 1 | 0.5 % |
| Ireland | 60 | 32.3 % |
| UK | 21 | 11.3 % |
| USA | 15 | 8.1 % |
| Libya | 36 | 19.4 % |
| UAE | 8 | 4.3 % |
| Saudi | 9 | 4.8 % |
| Malaysia | 10 | 5.4 % |
| Others | 23 | 12.7 % |

## 4.2. Education Level of Respondents

To complement the estimate of the depth of knowledge and understanding of respondents along with their age and year of work in the present organization, as discussed in previous sections, a question was inserted in a questionnaire asking, "what the highest degree is you have obtained?" Answers on education level of respondents are presented in Table 4.7. Responses constituted 35.5% respondents with master's degree, 34.9% of them with bachelor's degree, and 8.1% of total respondents had Ph.D. degree. Therefore, it can be concluded that the respondents had very good education levels, and their opinion and understanding are valuable for the analysis.

Table 3 Respondents' education level Constructs

| Education level | Frequency | Percent |
| --- | --- | --- |
| Highly National Diploma | 6 | 3.2 % |
| Diploma | 13 | 7.0 % |
| Secondary school | 10 | 5.4 % |
| Bachelor's degree | 65 | 34.9 % |
| Masters | 66 | 35.5 % |
| PhD | 15 | 8.1 % |
| Non | 5 | 2.7 % |

## 4.3. Measurement Model

All SEM related parameters were estimated by employing a maximum likelihood method in AMOS. The measurement model was tested by allowing all constructs' variance to vary freely. This is employed in Figure 5.1 by connecting the two-way arrow between all constructs. The measurement model is shown in Figure 5.1. As shown in the figure, each construct is connected through a bidirectional arrow. The overall measurement model fit is shown in Figure 5.1 below. The ratio of chi-square to the degree of freedom is 2.17 (1365.22/629) which is slightly more than 2 and therefore acceptable [43]. As presented in the Table 4. all fit indices, NFI, TLI, and CFI are close to 0.90 and RMSEA just greater than 0.06 are an indication that the study has an acceptable level of fit [7].

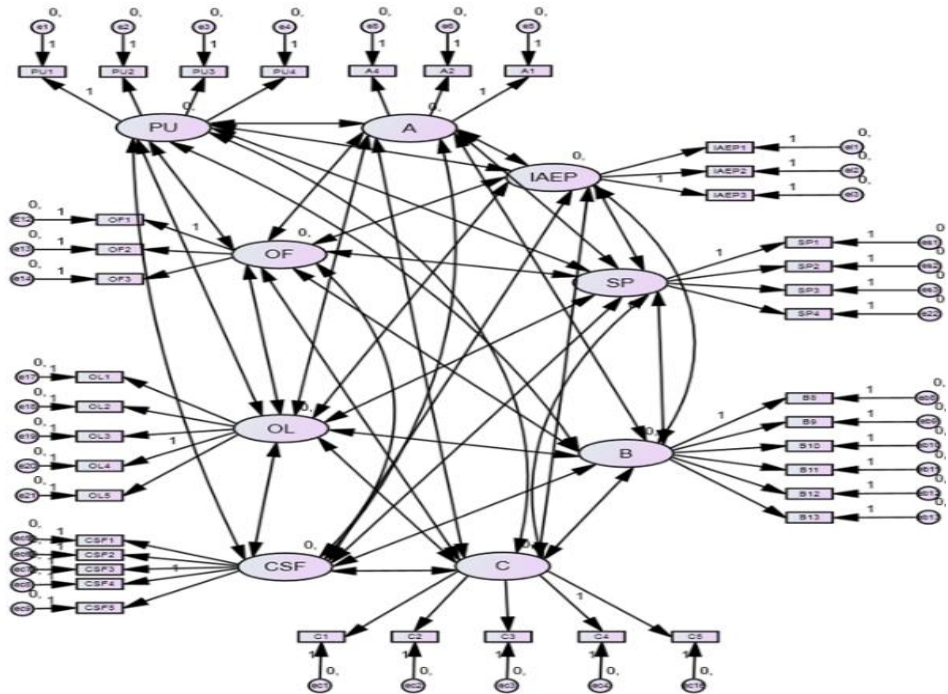

Fig.1 Measurement model

Table 4. Structural model fit indices

| Chi-square | Degree of freedom (DF) | p | NFI | CFI | TLI | RMSEA |
|---|---|---|---|---|---|---|
| 1365.21 | 629 | <0.001 | 0.808 | 0.885 | 0.872 | 0.08 |

Finally, the assessment of predictive validity which is the extent to which a measure predicts the criterion scores was performed. It can be assessed by examining the theory-driven relationship between various constructs which are hypothesized in the previous chapter. The structural model tests the relationships between hypothesized constructs. The relationship between constructs is assessed by examining sign (positive or negative), the value of path coefficient and its significance level. The regression (or path) coefficient of every path and its significance level are presented in Table

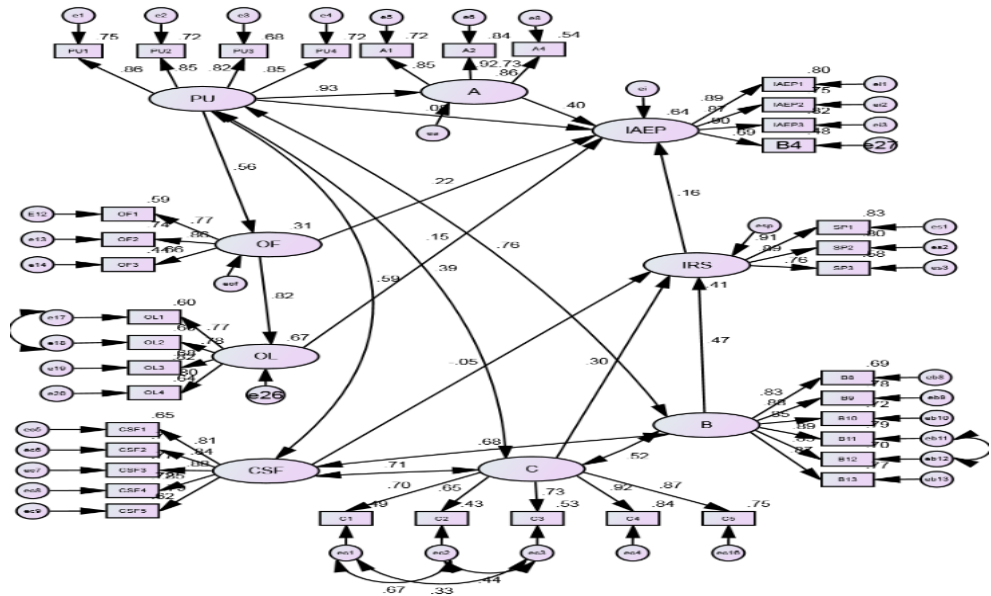

Fig.2 -Structural model

## 5. DISCUSSION

This study applies TAM theory in the context to adopt e-procurement and tries to integrate RIS with it [9]. A comprehensive model has been developed with the help of previous studies. To validate the model, the data collected from industries were analysed using confirmatory SEM techniques. The Results are presented in the previous section.

The newly designed E-PAM model is based on nine variables. The multiple regression analysis (MRA) using SPSS was performed to predict the value of final output (AEP) dependent upon the model variables (RIS, ATT, etc.). MRA is a standard technique to present the explanatory power and prediction of variable values [37]. Explanatory power of this study has R2 value of 64% to adopt e-procurement. These results are comparable to previous behavioural research studies [9], The path coefficient describes relationship between predictor and criterion construct. For

Example, if ATT increases by one standard deviation from its mean, AEP is expected to increase by 0.398 (see Table 5) its own standard deviation from its own mean while holding other regional relationships constant. As presented in the table, eight hypotheses are supported while there is a lack of evidence to supporting the other hypotheses. Four hypotheses are supported at p<0.001; two hypotheses are supported at p<0.01; one hypotheses was supported at p<0.06 and one hypotheses was supported at p<0.10.[4]. For example, in a similar study related to adoption of e-procurement using TAM, the observed variance for intention to use e-procurement was 62% [2]. By extending the TAM, another study was conducted in context of world-wide-web www usage [35]. The observed variance for actual usage of WWW context using that model was found to be 37.8% [2],, (Moon, J. and Kim, Y., 2001). Therefore, the overall explanatory power of the newly designed research model E-PAM was relatively higher for the adoption of e-procurement.

The variance for the variable attitude by using E-PAM was found to be 86%. This result for attitude variable is better than the previous model results for e-procurement (68%) and WWW context (33.2%) [2],.

| Hypothesis | Path coefficient (β) | Decision |
|---|---|---|
| H1: PU→AEP | 0.083 | Not supported |
| H2: ATT→AEP | 0.398* | Supported |
| H5: PU→A | 0.927*** | Supported |
| H6: OF→AEP | 0.222[a] | Supported |
| H7: PU→OF | 0.556*** | Supported |
| H8: OL→AEP | 0.151 | Not supported |
| H9: OF→OL | 0.816*** | Supported |
| H10: RIS→AEP | 0.16** | Supported |
| H11: CSF→RIS | -0.049 | Not supported |
| H12: CHB→RIS | 0.299** | Supported |
| H13: BNF→RIS | 0.472*** | Supported |

## 5.1. Non-Response Bias

It is important to know if there were significant differences between the respondents who responded soon after the online survey was distributed and those who responded after some time. As responses were collected online during of three months (13 weeks) respondents in the first 7 weeks of the study period were classified as early respondents and those from the final 6 weeks of the study were classified as late respondents. In total 100 participants were classified as early respondents and the majority of the responses in this category occurred in week 1 (30 responses) and week 4 (47 responses). A further 86 participants were classified as late respondents (week 8 to 13 inclusive) and a relatively smooth response pattern was observed during this period with the except on of the final week in which the number of participants was approximately double the number in any of the previous five weeks. The early and late responses were compared using chi-square test to confirm if the nonresponse bias occurred in this study. It was clear from the nonresponse bias results that only 13 of the 58 variables were significant, when investigated for early and late response. This indicated that non-response bias was not likely to occur in this study and, therefore, it is reasonable to conclude that the responses of the sample can safely be concluded to be representative of the larger population (respondents and non-respondents). This enables generalisation of the conclusions that will be generated from the statistical analysis of the sample responses to the survey.

Table 6. Significant Variables in Chi-square test for on-response bias

| Significant Variable | Value | p-value($p<0.05$) |
|---|---|---|
| OF3 | 186.0 | 0.045 |
| OF4 | 186.0 | 0.045 |
| OF5 | 186.0 | 0.045 |
| CHB1 | 24.397 | 0.018 |
| CSF2 | 21.054 | 0.050 |
| CSF3 | 23.941 | 0.047 |
| BNF 2 | 18.919 | 0.008 |
| BNF 4 | 17.890 | 0.036 |
| BNF 5 | 24.233 | 0.029 |
| BNF 6 | 19.457 | 0.022 |
| BNF 9 | 15.237 | 0.055 |
| BNF 12 | 16.784 | 0.019 |
| BNF 13 | 18.661 | 0.017 |

[2],[35] The observed variance for the organization facilitators and organizational leadership was found to be 31% and 66% respectively. The R2 value for Rule of Islamic Sharia variable was 41%. The values of R2 range from 31% to 86% representing a high level of explanatory power. This suggests that the newly designed model E-PAM, an extension of TAM, is capable of explaining a relatively high proportion of variance.

The result of the hypotheses H1, H2, H3, H4 and H10, which predicted positive relationship of PU, ATT, OF, OL, and RIS to AEP respectively, is presented in Table 5. The path coefficient (β) calculated by MRA for Hypothesis H2 (Attitude→AEP) was found to be 0.398 (p < 0.06). This result is comparable to previous study related to predicting e-procurement adoption in a developing country where TAM was used and the path coefficient for link between attitude and intention of e-procurement was 0.35 (with p<0.001) [2].

For H3 (OF→AEP, β = 0.222, p < 0.10), and H10 (RIS→AEP, β = 0.16, p < 0.01) are found significant and are supported by the data. However, hypothesis H1 (β = 0.083, p > 0.10) and H4 (β = 0.151, p > 0.10) could not find support from the data. Precisely, attitude, organisational facilitators, and Rules of Islamic Sharia are emerged as variables influencing positively the adoption of e-procurement. These hypotheses together explain a total of 64% (>50%) variance of AEP which is comparable to previous studies [2], (Moon, J. and Kim, Y., 2001). Going by B-coefficient, attitude (β = 0.39) appears to be the most important enabler for organisations to adopt e-procurement followed by organisational facilities (β = 0.22) and Rules of Islamic Sharia (β = 0.16). Attitude is predicted by perceived usefulness which explains 83% of the variance of the response variable. This relationship is found significant (β = 0.93, p < 0.001) and supported by the study. This result is comparable to previous study demonstrating path coefficient of 0.41 (p<0.001) between Usefulness and Attitude [2].

Regarding the direct link between perceived usefulness and adoption of e-procurement, it was observed in this study that the path coefficient between PU and AEP was 0.083. On the contrary, a previous study demonstrated a positive link between PU and intention to adopt internet banking with path coefficient 0.21 [28]. This finding was not expected from the data. One possible explanation for this finding is that due to tremendous use of online systems, social media, and other related applications, it has now become a common knowledge that information systems used for business processes helps in enhancing the effectiveness and is almost unavoidable. In this regard, it makes sense if organisations do not care much on perceived usefulness for implementing e-procurement.

The next hypotheses are related to perceived usefulness, organisational facilitators and leadership. Hypothesis predicting positive effect of perceived usefulness on organisational facilities is found significant (β = 0.55, p < 0.001) and supported by the study. The organisation facilitators has been considered important variable in models for adoption of information technology tools in previous studies, however investigation of direct linkage between OF and PU has never been performed using MRA as noted in the literature (Ellinger, A.D., Watkins, K.E. and Bostrom, R.P., 1999), [4]. The results from this doctoral study demonstrated that the perceived usefulness and organisational facilitators are positively linked with each other supporting in the adoption of e-procurement. When people recognise that there are ways which can increase working efficiency, ease in doing work with lesser skill sets, naturally they seem to acknowledge that there is a good number of organisation facilitators which reflects training for IT, training for e-procurement, and encouraging adoption of e-procurement. According to E-PAM, the Organisational Facilitators and Organisational leadership directly influence the e-procurement adoption as well as support each other. Both of these effects found significant and positive.

Therefore, the results of this study support the positive relationship of organisation facilitators to adopt e-procurement ($\beta$ = 0.22, p < 0.10). In addition to that, OF also support OL positively ($\beta$ = 0.82, p < 0.001). A large portion of variance of organisational leadership, 67%, is explained by organisational facilitators. This indicates organisation facilitators plays an important role in enhancing organisational leadership and is a positive predictor. Organisational leadership in this study reflects management support, lead role by management, invested time, effort & money, and be proactive in using e-procurement. All these measures of organisational leadership represent operational aspect of organisations. The better organisational facilitators could provide good leadership. This is also supported by the fact that a firm equipped with technologies, online systems, and machines will have a greater potential and incentive to go further in this direction and implement e-procurement. Further, the organisation facilitators impact adoption of e-procurement significantly and positively. This finding is in line with (Ellinger, A.D., Watkins, K.E. and Bostrom, R.P., 1999), [49] emphasised that IT sophistication (a deeper and wider penetration of IT) has positive impact on adoption of e-procurement. A moderate level of support was found for organisational determinants; however, the conceptual model was not tests using quantitative methods (Veit, D. J., Nils P. Parasie and Huntgeburth, J. C. 2011). (Ellinger, A.D., Watkins, K.E. and Bostrom, R.P., 1999) studied managers as facilitators of learning influences behaviours Therefore, the learning of an organisation (and employees) as a whole depend upon the behaviour of the managers (organisation leadership). However, organisational leadership which was postulated as impacting to adopt e-procurement positively, found insignificant in this doctoral study ($\beta$ = 0.15, p > 0.10), and therefore was not supported by the results. The actual cause of this unexpected result depicted by data needs to be investigated. However, one possible explanation of this result could be similar to the possible explanation of insignificant relationship of perceived usefulness to adoption of e-procurement. Adoption of e-procurement might be strongly driven by market which is increasingly automated and web-based across all spheres of business. Another cause could be the employees' increasing inclination towards accomplishing their jobs through automated processes and workflow. The natural inclination towards implementing e-procurement could have been a reason which neutralises organisation leadership playing a role in adopting e-procurement.

Another important set of hypotheses predicted a positive relationship of critical success factors, barriers & challenges, and benefits to Rules of Islamic Sharia. The critical success factors influencing Rules of Islamic Sharia is observed insignificant ($\beta$ = -0.05, p > 0.10). Barriers & challenges influencing Rules of Islamic Sharia is observed significant ($\beta$ = 0.30, p < 0.01).

Benefits influencing Rules of Islamic Sharia is observed significant ($\beta$ = 0.47, p < 0.001). The average variance calculated for RIS by correlation of CHB, CSF, BNF, and AEP was 0.80. The direct observed variance of RIS influencing AEP was 0.44. These results are satisfactory in the sense of positive influence of RIS on AEP as postulated in the beginning of this study.

Critical success factors may influence Rules of Islamic Sharia was postulated in the light of the study [47] highlighted influence of critical success factors in implementation perspectives and on the success of e-procurement. The implementation perspectives include organisation & management, practices & process, and systems & technology. However, this study finds a lack of support for confirming the positive influence of critical success factors on Rules of Islamic Sharia. Further, the data supported that barriers & challenges influence Rules of Islamic Sharia positively. This means that RIS positively mediates the barriers & challenges and adoption of e-procurement. The barriers & challenges reflect technological challenge, strategic challenge, human & process problems, misunderstanding of Islamic rules, and misunderstanding of non-Islamic rules. These barriers and challenges reflect the description of question 16 and 17 of online questionnaire.

When organisations go to implement any new practices and processes, it becomes apparent to them to evaluate barriers & challenges to implement the new practices. The findings from this doctoral study establishing positive relationship between RIS and CHB and could encourage the RIS observing companies and groups to adopt e-procurement. It was found that the benefits impact Rules of Islamic Sharia positively. The benefits include improved communication and collaboration with suppliers, decentralise procurement management, reducing costs, allowed the purchasing department to concentrate on more strategic tasks, reduction of purchasing department size and number of functional areas, and improved effectiveness of purchasing process. This positive relationship between BNF and RIS conveys that more the benefits derived from adopting e-procurement, organisations would go for Rules of Islamic Sharia. This notion has been approved from the results especially for those organisations operating under Rules of Islamic Sharia or bound to operate under Sharia rules because when organisations get enough benefits from e-procurement, there would be an incentive to experiment what-if by falling in the line of Sharia rules. In case of compulsion for firms to operate under Rules of Islamic Sharia, organisations may feel more comfortable to follow Sharia rules when benefits would be more [32].

The hypothesis which postulated a positive relationship between Rules of Islamic Sharia and to adopt e-procurement is found significant ($\beta = 0.16$, $p < 0.01$) and thus supported. These findings imply that organisations operating under RIS may find it beneficial to adopt e-procurement. Therefore, this study suggests managers of organisations operating under Rules of Islamic Sharia to go for e-procurement rather than functioning with paper-based procurement. This switching to e-procurement would be promising.

## 6. CONCLUSION AND FUTURE RESEARCH

In this research, several studies about off to the adoption of e-procurement have been reviewed, and based on this knowledge a conceptual model has been built. To test the conceptual model, a questionnaire was developed, and an online survey has been conducted. The conceptual model was analyzed by using confirmatory SEM techniques on the data collected. Overall, it was found that attitude, organisation facilities, and Rules of Islamic Sharia are important in influencing the adoption of e-procurement strategy. These factors together explain a highly significant 64% of the variance of adoption of e-procurement. Barriers & challenges and benefits are found to positively impact upon the Rules of Islamic Sharia mediator. Contrary to the many existing studies, not enough evidence was found to support the hypothesis that perceived

Usefulness positively impacts the adoption of e-procurement. The possible reason for this has been explained. Therefore, it is good for organizations to focus on attitude, organizational facilitators and Rules of Islamic Sharia. Support was also found for perceived usefulness influencing organizational facilitators and organization facilitators influencing organizational leadership positively. Although, perceived usefulness was not observed to influence the adoption of e-procurement directly, it has been identified as impacting the adoption of e-procurement through attitude and organizational facilitators. Therefore, the role of perceived usefulness to adopt e-procurement cannot be ignored; rather it needs to be strengthened actively.

The hypothesis which assumed a positive relationship between the Rules of Islamic Sharia and an aim to adopt e-procurement is proved significant and therefore sustained and supported. These findings imply that organizations which operate under Sharia law should find it beneficial to adopt e-procurement technology. Therefore, this study recommends that managers of organizations operating under Rules of Islamic Sharia adopt e-procurement rather functioning with paper-based procurement.

E-procurement in developing countries is still in its infancy stage; therefore future studies of this subject may encounter different factors resulting from the rapid change in technology and other issues that may not have been observed in this research.

The conceptual model that has been developed in this research is to help the developing countries in the Islamic region to adopt and implement e-procurement strategy. Therefore, a study collecting data only from the Islamic region and evaluating the research model would be beneficial.

Future studies could explore and examine new factors that this research has not investigated in detail. Future researches could also focus on certain sectors in the Islamic context and separate them to from research areas to be able to understand in depth the challenges and barriers facing such adoption of technology and the benefits and success factors from such adoption.

To improve in-depth understanding and reasons for all significant and insignificant relationships, future research might consider a qualitative approach such as grounding theory or case studies.
The longitudinal study can be used to understand if the model and its underlying theories vary over different time periods.